\documentclass[aps,prl,reprint,superscriptaddress,showpacs]{revtex4-1}
\usepackage[latin9]{inputenc}
\usepackage{color}
\usepackage{amsmath}
\usepackage{amssymb}
\usepackage{graphicx}

\makeatletter
\usepackage{eurosym}
\makeatother

\begin{document}

\title{On the Ohm law in dilute colloidal polyelectrolytes}

\author{I.~Chikina}
\affiliation{IRAMIS, LIONS, UMR NIMBE 3299 CEA-CNRS, CEA-Saclay, F-91191 Gif-sur-Yvette
	Cedex, France}
\author{V.~Shikin}
\affiliation{ISSP, RAS, Chernogolovka, Moscow District, 142432 Russia}
\author{Andrey Varlamov}
\affiliation{CNR-SPIN, c/o DICII-Universit{á} di Roma Tor Vergata, Via del Politecnico, 1, 00133 Roma, Italy}

\date{\today }
\begin{abstract}
We discuss the peculiarities of the Ohm law in dilute polyelectrolytes
containing a relatively low concentration $n_{\odot}$ of the multiply-charged
colloidal particles. It is demonstrated that in this conditions, the
effective conductivity of polyelectrolyte is the linear function of
$n_{\odot}$. This happens due to the change of electric field in
the polyelectrolyte under the effect of colloidal particle polarization.
 Such mechanism gives grounds to propose the alternative scenario 
for the phenomenon observed experimentally.
\end{abstract}

\maketitle

\section{Introduction}

In the Refs.~\cite{Lucas,Saco1}, among various observations,
the authors report the effect of multiply-charged colloidal particles
(with the effective charge $eZ\gg e$) on conductivity of the dilute
polyelectrolytes. It turns out that the latter grows linearly with
increase of the colloidal particles concentration. This
finding seems to be non-trivial from different points of view, and
in the first hand in the optics of the percolation theory
(see, for example, \cite{Shklovski}). Indeed, in accordance to the
percolation theory the conductivity of a mixture between dielectric
and conducting components, remains minute until the fraction of conducting
phase approaches the percolation threshold. And only in vicinity of
the latter the conductivity growths smoothly from the value of the
dielectric component to that of metallic one.

Before discussing this contradiction, let us make an excursus into
the physics of semiconductors. In the theory of semiconductors \cite{Shklovski},
the regions of weak and strong doping (i.e., introduction of charged
impurities or structural defects with the purpose of changing the
electrical properties of a semiconductor) are distinguished. In the
low doping regime, the impurity concentration $n_{\odot}$ is so small
that the distances between them significantly exceed the Debye length
$\lambda_{0}$ and the bare radius of the colloidal particle $R_{0}$, i.e.
\begin{equation}
n_{\odot}\left(\lambda_{0}+R_{0}\right)^{3}\ll1,R_{0}\leq\lambda_{0},\label{criterion}
\end{equation}
and the intrinsic charge carriers of semiconductor completely screen
the electric fields produced by the charged impurities (see Fig. 1). In the strong
doping regime, when the criterion (1) is violated, the fields produced
by the dopants are screened only partially and their interaction becomes
significant.

Returning to the case of the dilute colloidal polyelectrolytes one
can identify $n_{\odot}$ with the concentration of the colloidal particles, while $\lambda_{0}$ is related to their characteristic size. The latter
is determined by the known concentration $n_{0}$ of the counterions
of the electrolyte hosting the charged colloidal particles.

The criterion (\ref{criterion}) is in a reasonable agreement to the
common concepts of the physics of dilute polyelectrolytes developed
in Ref. \cite{Derjaguin,Verwey,LLSP} and known as DLVO formalism.
Namely, if the colloidal particles are neutral, they can not exist
stationary in dilute solution, coagulating due to the van der Waals
forces acting between them. In order to prevent such coagulation processes,
one can immerse individual colloidal particles in the electrolyte
specific for each sort of them. The latter are called stabilizing
electrolytes.

Being immersed (or synthesized within) an electrolyte solution, the nanoparticles acquire surface ions (e.g., hydroxyl groups, citrate, etc. \cite{Riedl,Bacri,Dubois}) resulting in a very large structural charge $eZ$ ($|Z|\gg10$). Its sign can be both positive and negative, depending on the surface group type. The latter in return, attracts counterions from the surrounding solvent creating an electrostatic shielding coat of the size $\lambda_0$ with an effective charge $-eZ$.  In these conditions, nano-particles approaching between them to the distances $r\le\lambda_{0}$ begin to repel each other without floculation \cite{Derjaguin,LLSP,Verwey}.
The region of an essential interaction between them in terms of the criterion (\ref{criterion}) corresponds to the condition 
\begin{equation}
n_{\odot}^{c}\left(\lambda_{0}+R_{0}\right)^{3}\sim1.\label{ncr}
\end{equation}

Coming back to the results of  Ref.~\cite{Lucas,Saco1} one can
see that observation of the linear growth of the dilute colloidal
polyelectrolytes conductivity $\sigma$ as the function of the concentration
$n_{\odot}$ is in the evident contradiction with the percolation
theory \cite{Shklovski}. Indeed, each massive multiply-charged colloidal
particle is surrounded by the cloud of counter-ions screening its
positive charge. Such formations, according to Ref. \cite{Shklovski},
should not affect on conductivity of the dilute solution until shells
of the neighbor charged complexes will not overlap among themselves
(see Eq. (\ref{ncr})). The results of Ref.~\cite{Saco1} demonstrate the
opposite: the conductivity of dilute colloidal polyelectrolyte grows
linearly with increase of concentration already in the range $n_{\odot}\ll n_{\odot}^{c}$,
where there is not yet place for percolation effects.

\begin{figure}[!ht]
	\includegraphics[width=1\linewidth]{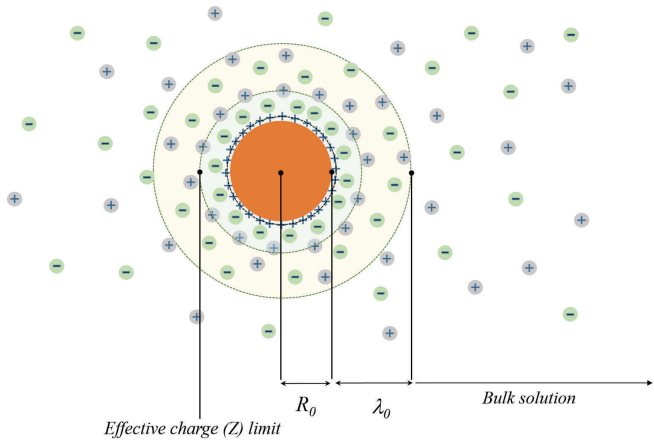} \caption{The schematic presentation of the multiply-charged colloidal	particle  surrounded by the cloud of counter-ions.}
	\label{Figure_1} 
\end{figure}

This contradiction can be eliminated by noticing that presence of
the multiply-charged colloidal particles has an effect not only on
the value of conductivity of solution but also on the local value
of the electric field: 
\begin{equation}
j(n_{\odot})=\sigma(n_{\odot})E(n_{\odot}).\label{jnonlin}
\end{equation}
Important to note that the factors in Eq. (\ref{jnonlin}) are affected
by presence of the multiply-charged colloidal particles in different
ways. While the conductivity of the electrolyte at low concentrations
of multiply charged colloidal particles ($n_{\odot}\leq n_{\odot}^{c}$)
remains almost unchanged, their effect on local electric field in
this range of concentrations is essential. This happens due to polarization
of the colloidal particles by an external electric field which, in
accordance to the Le Chatelier's principle, results in the decrease
of the effective value of the field. Consequently, the growth of conductivity
~\cite{Lucas,Saco1} as the function of concentration $n_{\odot}$
is observed in experiment. When the concentration of multiply charged
colloidal particles reaches the percolation threshold ($n_{\odot}=n_{\odot}^{c}$)
the role of factors in Eq. (\ref{jnonlin}) is reversed. Here the
subsystem of colloidal particles forms clusters and cannot be considered
more as the gas of polarized highly conducting particles. Yet, namely
in this range of concentrations, the new channel of percolation charge
transfer is opened and the total conductivity of the electrolyte growth
with further increase of $n_{\odot}$.

The state of art of transport phenomena in polyelectrolytes recently
was reviewed in Ref.~\cite{Lucas}. Focusing mainly on the results
of the microscopic approach \cite{Dufreche,Durand,Jardat} the authors
discuss in details  mobility, diffusion coefficient, effective
charge space distribution of the colloidal particles as the function
of their concentration. Yet, in Ref. \cite{Lucas} there is no any
information concerning the effect of clusters polarization on the
charge transfer process in such complex systems. Namely this aspect
of the problem is the subject of our work.

\section{Effective electric field in bulk of colloidal polyelectrolyte}

The colloidal polyelectrolyte presents itself a weakly conducting
liquid with the small but finite fraction of relatively highly (due
to $Z\gg1$ ) conducting inclusions: colloidal particles. The collective
polarization of these inclusions occurs when the external electric
field $E_{0}$ is applied. This phenomenon is analogous to polarization
of neutral atoms in gas. The only difference is that the neutral atoms
reside in vacuum, while the charged conducting clusters of colloidal
polyelectrolyte are immersed in a less, but still conducting, medium.
Hence our goal is to account for this peculiarity and find the effective
field which governs the charge transport in such complex system.

\subsection{Electric field in absence of current}

The space distribution of the effective electric field of the colloidal
particle is determined by the Poisson equation (see \cite{Shklovski,LLSP})
\begin{equation}
\Delta\varphi=\frac{4\pi}{\epsilon}\rho(r),\quad\rho(r)=|e|[n_{+}(r)-n_{-}(r)],
\label{poisson}
\end{equation}
where $\epsilon$ is the dielectric permittivity of stabilizing electrolyte. 

The concentrations of the screening counterions $n_{\pm}(r)$ is determined
self-consistently via the value of local electrostatic potential 
\begin{equation}
n_{\pm}(r)=n_{0}\exp{[e_{\pm}\varphi(r)/T]},
\end{equation}
 $n_{0}=n_{0}^{+}=n_{0}^{-}$ is the counterions bare concentration,
occuring due to the complete dissociation of the electrolyte which
stabilizes the gas of colloidal particles.

In assumption $e\varphi(r)<T$ the Poisson equation can be linearized
and takes form
\begin{equation}
\Delta\varphi=\varphi/\lambda_{0}^{2},\quad\lambda_{0}^{-2}=\frac{8\pi e^{2}}{\epsilon T}n_{0}.
\end{equation}
This equation should be solved accounting for the boundary conditions
\begin{equation}
r\varphi(r)_{|r\to R_{0}}\to Z|e|,\quad\varphi(r)_{|r\to\infty}\to0,
\end{equation}
what results in the standard screened Coulomb potential:
\begin{equation}
\varphi(r)=Ze\frac{\exp(-\frac{r}{\lambda_{0}})}{r}.
\label{Yukava}
\end{equation}
The values $Z$,$R_{0}$ and $n_{0}$ of the electrolyte,
which stabilizes the colloidal solution can be determined by independent
experiments (for example, by measurements of the electrophoretic forces,
osmotic pressure, etc. \cite{Lucas}). 

One should remember that even strongly diluted polyelectrolytes can
undergo the transition to the state of Wigner crystal in the case
of strongly charged colloidal particles ($Z\gg1$). \textcolor{black}{For
description of this, observed experimentally }\cite{heltner,Kose,Williams},\textcolor{black}{{}
phenomenon the authors of }\cite{Alexander} assumed that the interaction
between two colloidal particles has the same form of Yukawa
potential (8), yet with the renormalized effective charge \textcolor{black}{$Z^{*}\ll Z,$
explicitly depending on the colloidal particles density $n_{\odot}$.
The value of $Z^{*}$ is determined in the Wigner-Seitz model from
the new boundary condition 
\[
\frac{\partial\varphi}{\partial r}|_{r\rightarrow n_{\odot}^{-1/3}}=0
\]
replacing that ones, valid for the isolated charged particle in the
screening media (see Eq. (7)). For some range of the colloidal particles
densities $n_{\odot}$ the conditions} $Z\gg1$ and
$Z^{*}\ll1$ can be satisfied simultaneously. The former characterizes
the properties of the multiply-charged colloidal particles itself,
while the latter is determined by strength of their interaction and
$n_{\odot}$. In the range of densities
$n_{\odot}$ satisfying Eq. (1) the effect of the effective charge
$Z^{*}$ on the Ohmic transport is negligible.

\subsection{Electric field in presence of current}

When a stationary current flows through the polyelectrolyte, an internal
electric field $\vec{E}$ appears in it. In the approximation of a
very diluted solution one can start consideration from the effect
of presence of the isolated colloidal particle on flowing current.
Namely, one should find the perturbation of the internal electric
field which would provide the homogeneity of the transport current
far from the colloidal particle. Corresponding problem recalls that
one of the classic hydrodynamics: calculus of the associated mass
of the particle moving in the ideal liquid \cite{LLH}. 

We choose the center of spherical coordinates coinciding with the
colloidal particle and direct the $z-$axis along the electric field
$\vec{E_{0}}$. We assume that the conductivity of the electrolyte
in absence of colloidal particles is $\sigma_{0}$. The highly charged colloidal particle we will model as the conducting solid sphere of the radius $R \simeq (R_0 + \lambda_0)  $(see Fig 1) with
conductivity $\sigma_{\odot}>\sigma_{0}$.  Analysis of the charge
transport in multi-phase systems (see \cite{Jackson}) is based on
the requirements
\begin{equation}
div\vec{j}=0,\quad\vec{j}=\sigma\vec{E}. \label{cont}
\end{equation}
When the medium conductivity is invariable in space the constancy
of current automatically means the homogeneity of the electric field.
The situation changes when the system is inhomogeneous and $\sigma\ne const$.
The continuity equation (\ref{cont}) in this case should be solved with the
boundary conditions accounting for the current flow through the boundaries
between domains of diverse conductivity. According to Ref. \cite{Jackson,Dykhne}
the tangential components of electric field intensity at the boundary
must be continuous, while the normal ones provide the continuity of
the charge transfer. Applying these rules to our simple model of the
 highly charged colloidal particle in the less conductive
medium one can write
\begin{equation}
j_{n}^{0}=J_{n}^{\odot},\quad\mbox{or}\quad\sigma_{0}E_{0}=\sigma_{\odot}E_{\odot}. \label{bound}
\end{equation}

Solution of the system of Eqs. (\ref{cont}),  (\ref{bound}) for the electrostatic
potential in the vicinity of the colloidal particle ($r\ge R$) acquires
the form:
\begin{equation}
\varphi(r,\theta)=-E_{0}r\cos{\theta}+\left(\frac{\gamma-1}{\gamma+2}\right)E_{0}\frac{R^{3}}{r^{2}}\cos{\theta},\label{phitheta1}
\end{equation}
with $\gamma=\sigma_{\odot}/\sigma_{0}$. In the limit $\gamma\rightarrow1$
the electric field remains unperturbed, $\vec{E}=-\nabla\varphi\to\vec{E}_{0}$.
In the opposite case, $\gamma>>1$, the dipole perturbation takes
the form corresponding to the case of metallic inclusion of the radius
$R$ in the weakly conducting environment (Ref. \cite{Jackson}):

\begin{equation}
\varphi(r,\theta)=-E_{0}r\cos{\theta}\left(1-\frac{R^{3}}{r^{3}}\right).\label{phitheta2}
\end{equation}

One can see that in accordance to the intuitive expectations presence
of an isolated colloidal particle in electrolyte leads to appearance
of the local perturbation of the electric field of the dipole type
$\nabla\varphi\propto r^{-3}$ with the value of the dipole moment
of one colloidal particle

\begin{equation}
p_{\odot}=\left(\frac{\gamma-1}{\gamma+2}\right)R^{3}E_{0}.\label{psimple}
\end{equation}

Returning to the initial problem of the rarefied gas of colloidal
particles of the concentration $n_{\odot}$ in the electrolyte media
one can introduce the effective dielectric permittivity $\epsilon_{\odot}$.
It can be related to the dipole moment (\ref{psimple}) by means of the Clausius--Mossotti
relation (see Ref. \cite{Jackson}) and in terms of the material parameters
of the problem is read as:
\begin{equation}
\epsilon_{\odot}=1+4\pi\left(\frac{\gamma-1}{\gamma+2}\right)R^{3}n_{\odot}. \label{epsilon}
\end{equation}

One can try to make the model of colloidal particle more realistic
assuming that the latter has the structure of thick-walled sphere:
a ``nut'' with the conducting shell and the insulating core of the
bare radius $R_{0}$. This intricacy leads to the change in the expression
for the corresponding dipole momentum: instead of Eq. (13) it takes
form (see Ref. \cite{Jackson})

\begin{equation}
\tilde{p}_{\odot}=\frac{\left(2\gamma+1\right)\left(\gamma-1\right)}{\left(2\gamma+1\right)\left(\gamma+2\right)-2\left(\gamma-1\right)^{2}R_{0}^{3}/R^{3}}\left(R^{3}-R_{0}^{3}\right)E_{0}.\label{eq:pcomplex}
\end{equation}

This formula contains two geometrical parameters: $R$ and $R_{0}$.
The latter should be determined from some independent measurements.
 The difference $R-R_{0}$ one can identify with the Debye length $\lambda_{0}$ or to consider it as
the fitting parameter.

\section{Ohmic transport in a weak colloidal polyelectrolyte }

Eq. (\ref{epsilon}) demonstrates that growth of the nano-particles concentration $n_{\odot}$
leads to increase of the dielectric constant $\epsilon_{\odot}$,
what, in its turn, results in the decrease of the effective electric
field in electrolyte. The latter, in conditions of the fixed transport
current, is perceived as the growth of conductivity with increase
of the colloidal particles concentration:
\begin{equation}
\sigma(n_{\odot})\!=\!j\epsilon_{\odot}/E_{0}\!=\!\sigma_{0}\!\left[1\!+\!4\pi n_{\odot}\frac{p_{\odot}\left(E_{0}\right)}{E_{0}}\right].\label{eq:sigmasimple}
\end{equation}
This expression can be already used for the experimental data processing. 

\begin{figure}[!ht]
	\includegraphics[width=1\linewidth]{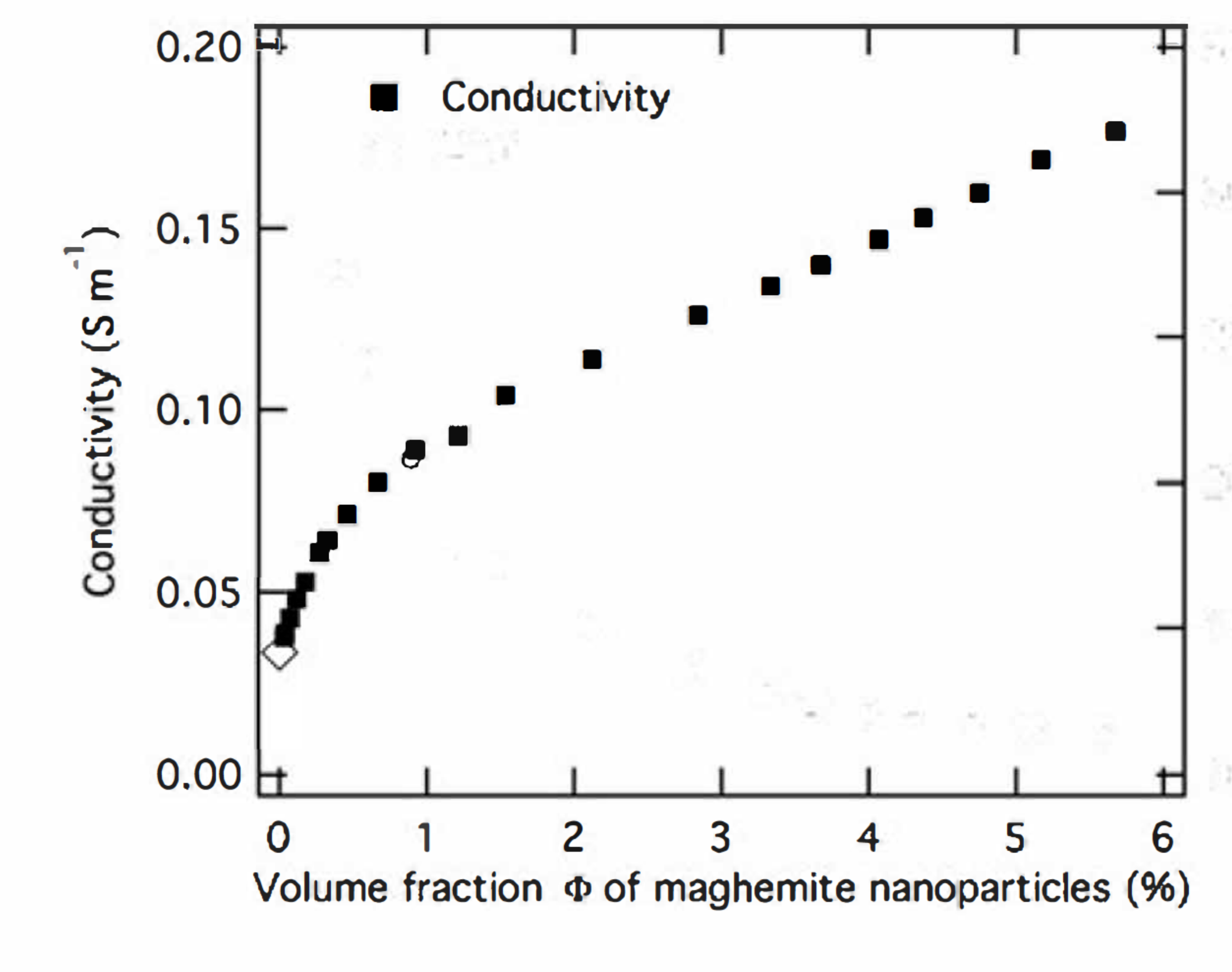} \caption{Experimental values of electrical conductivity of water based polyelectrolyte solution as a function of colloidal concentrations.   Measurements were performed on pH = 3.1 solutions containing maghemite nanoparticles with an average diameter of 12 nm.  More detailed information on the colloidal solution preparation methods and the nature of other ions are found in Ref.\cite{Lucas,Saco1}.}
	\label{saco_conductivity_corr} 
\end{figure}

\subsection*{Approximation of the conducting spheres}

Substituting the dipole moment taken in the approximation of Eq.~(\ref{psimple}) in Eq.~(\ref{eq:sigmasimple}) one finds 
\begin{equation}
\frac{\Delta\sigma_{\mathrm{CP}}(n_{\odot})}{\sigma_{0}}=\frac{\sigma(n_{\odot})-\sigma_{0}}{\sigma_{0}}=4\pi n_{\odot}\left(\frac{\gamma-1}{\gamma+2}\right)R^{3},\label{eq:exp1}
\end{equation}
where  $\Delta\sigma_{\mathrm{CP}}$ is the excess conductivity due to the presence of colloidal particles. The left-hand-side of this equation can be extracted from the data
presented in Fig. 2. Indeed, in the interval of the nanoparticles concentrations
$0 \le \phi \le 0.6\%$ the behavior of conductivity $\sigma(n_{\odot})$ is
almost linear and $\sigma(n_{\odot})/\sigma_{0}-1=0.7$. In its turn the
concentration $\varphi=0.6%\ensuremath{\text{\%}}
$\% corresponds to $n_{\odot}^{(1)}=5.45\centerdot10^{15}cm^{-3}$.

For the further estimations will be  crucial that the Eq. (\ref{eq:exp1}) is sensitive to the value of $\gamma$ only when it is not very large. When  $\gamma>>1$ (we will justify this limit below) the combination $(\gamma -1)/(\gamma +2) \rightarrow 1$ and it ceases to influence on the evaluations based on Eq.~(\ref{eq:exp1}).  This allows us to find in this limit 
\begin{eqnarray}
 R_{exp}^{(1)}=2.17\centerdot10^{-6}cm,  \nonumber \\ n_{\odot}^{(1)}\left[R_{exp}^{(1)}\right]^{3}=0.055\ll1.
 \label{evaluations}
\end{eqnarray}
One can see that these values, together with the nanoparticles
concentration $n_{\odot}^{(1)}$, confirms validity of the assumed above approximation (\ref{criterion}). The plausible reasons for the discovered considerable difference between $R_{exp}^{(1)}$ and  the value of bare radius $R_0^{(1)}=6\centerdot10^{-7}cm$ given in Ref. \cite{Lucas} we will discuss below. 

The above  found  conductivity correction $\Delta\sigma_{\mathrm{CP}}(n_{\odot}) \propto n_{\odot} R^{3}$ (see Eq. (\ref{eq:exp1})) caused by presence of nanoparticles in electrolyte  can be confidently distinguished from the standard Onsager-Debye conductivity ($\sigma_{\mathrm{OD}}$) of the diluted 1:1 electrolyte \cite{O27,DH,LLPK}.  Indeed, first of all the concentration dependencies of these conductivities are different:  $\Delta\sigma_{\mathrm{CP}}(n_{\odot}) \propto n_{\odot}$ while $\sigma_{\mathrm{OD}}(n_{\odot}) \propto \sqrt{n_{\odot}}$ .

Let us attract attention at the unusual dependence of the excess conductivity (\ref{eq:exp1}) on the nanoparticle size:  $\Delta\sigma_{\mathrm{CP}}$ growths with increase of $R$. Usually, this dependence is supposed to be opposite  (the larger radius of the sphere in Stokes viscous law, the lower its mobility, and hence, the conductivity).

One can analyze the available experimental data on conductivity of the stabilized diluted colloidal solution \cite{Lucas,Saco1}  in the conditions described by Eq. (\ref{ncr}).  In accordance to  Eq. (\ref{eq:exp1}) the excess conductivities for different sizes of nanoparticles in assumption of the same concentration should scale as $[R_{0}^{(1)}/R_{0}^{(2)}]^3$.  Taking  the value $R_{0}^{(1)}=6 nm$ from \cite{Lucas} and $R_{0}^{(2)}=3.8 nm$ from \cite{Saco1} one finds that the ratio
\begin{equation}
\frac{ \Delta \sigma_{\mathrm{CP}}^{(1)}}{\sigma_{0}^{(1)}}/
	\frac{\Delta\sigma_{\mathrm{CP}}^{(2)}}{\sigma_{0}^{(2)}}=\left(\frac{6}{3.8}\right)^3\approx 4
	\label{eq:exp3}
\end{equation}
Experimental data for this value give even more striking difference:
\begin{equation}
\frac{ \Delta \sigma_{\mathrm{CP}}^{(1)}}{\sigma_{0}^{(1)}}/
\frac{\Delta\sigma_{\mathrm{CP}}^{(2)}}{\sigma_{0}^{(2)}}=\frac{0.7}{0.06}\approx 11.7.
\label{eq:exp4}
\end{equation}

\subsection*{Approximation of the conducting thick-walled spheres}

Here it is necessary to notice that the value $R_{exp}^{(1)}$ obtained
in the simple approximation of Eqs. (\ref{psimple}), (\ref{eq:sigmasimple})
and the measured in Ref. \cite{Lucas} bare radius of the colloidal
particle $R_0$ form relatively small numerical parameter $[R_{0}/R_{exp}^{(1)}]^{3} \simeq 0.02 $.  It is why makes sense to improve
the experimental data proceeding replacing the value $p_{\odot}$
in Eq.~(\ref{eq:sigmasimple}) by the two parametric expression (\ref{eq:pcomplex}).
Tending $\gamma\rightarrow\infty$ in it one finds
\begin{equation}
\frac{\sigma(n_{\odot})-\sigma_{0}}{\sigma_{0}}=4\pi n_{\odot}[R_{exp}^{(1)}]^{3}\left[1-\frac{3}{\gamma}-\frac{9}{2\gamma}\frac{R_{0}^{3}}{\left([R_{exp}^{(1)}]^{3}-R_{0}^{3}\right)}\right]\label{eq:sigmasimple-1-1-1-1}
\end{equation}
From this expression  it is clear that the approximation (\ref{eq:exp1}) is valid when $\gamma \gg 1$.

The  parameter $ \gamma $ requires a special discussion.  In the DLVO colloidal model  it is assumed that some bare core exists which is able to cause the van der Waals forces between colloidal particles in dilute, non-stabilizing  solution. What are the conducting properties  of this core is not so essential.  For example, one can suppose this bare core of the radius $R_{0}$ to be semiconductor possessing its intrinsic charge carriers which are confined in its volume. If the solvent possesses the stabilizing properties its own mobile charge carriers, counterions, have the same properties, as the  intrinsic charge carriers of the bare core. The requirement of electrochemical potential constancy leads to the charge exchange between the bare core and the solvent. Such exchange results in the formation of the Debye shell (see Eqs. (\ref{poisson})-(\ref{Yukava}), where the concentration of counterions considerably exceeds that one in the solvent bulk.  We assumed above that the value of corresponding conductivity $\sigma(n_{\odot})$ considerably exceeds that one  $\sigma_{0}$  of the electrolyte  conductivity in absence of the nanoparticles. This assumption ($ \gamma \gg 1 $) breaks when  the average value of electrochemical potential in the Debye shell  $e\phi_{\odot}$, exceeds  temperature. The authors of  Ref. \cite{GNS} state that in these conditions the Debye shell of the DLVO colloid can crystallize due to Coulomb forces and the latter becomes insulator with $\sigma(n_{\odot}) \lesssim \sigma_0$.

\section{Conclusions}

The main result of this work consists of the proposition
of  alternative scenario explaining the linear growth of the polyelectrolyte
conductivity versus the concentration of colloidal particles observed
in Ref. \cite{Lucas,Saco1} in the conditions of the validity of Eq.~(\ref{criterion}).
It drastically differs from the existing ideas of the transport in
electrolytes resulting in the empirical Kohlrausch's law (see \cite{RS,LLPK})
\begin{equation}
\Delta\sigma\sim\sqrt{n_{\odot}}.\label{eq:calrush}
\end{equation}
The speculations justifying Eq.~(\ref{eq:calrush}) were firstly proposed
in so early papers as Ref. \cite{DH,O27} and the recent efforts
to improve this mechanism were undertaken in the Ref. \cite{LG13}. 

The fact of observation of the Ohmic transport in strong electrolyte
(Ref.~\cite{Lucas,Saco1}) denies the applicability
of the Kohlrausch's law in the interval of very low concentration of the
colloidal particles. Conversely, the mechanism proposed above,
based on the analogy to percolation
mechanism of conductivity occurring in doped semiconductors, allows
to get an excellent agreement  in the observed linear dependence.
Moreover, it also provides very reasonable values of the microscopic parameters
of the problem. 

One can believe that the validity  Kohlrausch's  law is
restored in the domain of higher concentrations and the crossover
point between the two regimes (\ref{eq:sigmasimple}) and (\ref{eq:calrush})
is determined by the condition (\ref{ncr}), as it is shown in Fig.
2. One can find the pro-arguments for this statement also in the experimental
curve shown in Fig. 2 of Ref. \cite{Lucas}, where the regimes are changed in  vicinity
of the concentration $n_{\odot}^{(1)}=5.45\centerdot10^{15}cm^{-3}$.

%Let us stress that remaining in the frameworks of our scenario of the %Ohmic transport accounting for the polarization of colloidal particles we %have succeeded to explain
%the main experimental finding ignoring the electrophoretic motion of
%the nanoparticles at all.

The natural question arises: why such linear growth
below the percolation threshold was never reported in measurements
performed on semiconductors? The answer probably consists in the overwhelming
supremacy of the colloidal particle dipole momentum in comparison
to that one of the dopant in semiconductor. 

It would be interesting to compare the values of effective charge $Z$ extracted from the experiments on conductivity of Ref. \cite{Saco1}  and the review article \cite{Lucas}. 
Unfortunately, this is not easy to do since analysis of the data for different $Z$ results in very different values of $R_0$. It is why one cannot judge about the influence of the effective charge $Z$ on the bare radius of the colloidal particle $R_{0}$.

The relative insensibility of the polyelectrolyte conductivity on the value of the parameter $Z$. is not extended on Seebeck coefficient. The measurements of Ref. \cite{Saco1}  demonstrated existence in its kinetics of the two different phases: the initial and steady ones. The authors dealt with two types of colloids, one is almost electroneutral ($Z\geq 1$.), the other is supposed to have $Z\gg 1$.

%The initial phase of the  effect in the polyelectrolytes with both kinds of the colloidal particles are similar, while the final steady phase differs strongly.  %The thermoelectric properties of the polyelectrolyte with weakly charged colloidal particles do not change much, the temperature gradient results only in %formation of inhomogeneity in their density (Soret effect).  Contrary, when temperature gradient is applied to the polyelectrolyte containing strongly %charged nanoparticles, the situation changes. The process of colloidal nucleus screening is transformed by the induced space inhomogeneity: the %dipoles start to move, what has to be taken into account in study of the Seebeck signal of such complex system. This will be done elsewhere. 

\section{Acknowledgments} 

The authors acknowledge multiple and useful discussions with Dr. Sawako Nakamae. This work is supported by  European Union's Horizon 2020  research and innovation program under the grant agreement n 731976 (MAGENTA).


\begin{thebibliography}{10}
\bibitem{Lucas} I.~Lucas, S.~Durand-Vidal, O.~Bernard, et al,
 \textit{``Influence of the volume fraction on the electrokinetic properties of maghemite nanoparticles in suspension''}, \textit{Molecular Phys.}, \textbf{112}, 112, (2014). 

\bibitem{Saco1} Thomas J.~Salez, et al, \textit{``Can charged colloidal
particles increase the thermoelectric energy conversion efficiency?}
\textit{Phys.Chem.Chem.Phys.}, \textbf{19}, 9409 (2017).

\bibitem{Shklovski} B.I.~Shklovski, A.L.~Efros, \textit{``Electronic
properties of doped semiconductors''}, Springer-Verlag, Berlin-Heidelberg-New
York-Tokyo (1984). 

\bibitem{Derjaguin} B.V.~Derjaguin L.D.~Landau, \textit{``Theory
of the stability of strongly charged lyophobic sols and of the adhesion
of strongly charged particles in solutions of electrolytes''}, \textit{Acta
Physico Chemica URSS}, \textbf{14}, 633 (1941).

\bibitem{LLSP} L.D.~Landau, E.M.~Lifshitz, \textit{``Statistical
Physics''}, (Volume 5 of a Course of Theoretical Physics),  3rd Edition, Elsevier, (2011). 

\bibitem{Riedl} J. C. Riedl, M. A. Akhavan Kazemi, F. Cousin, E. Dubois, S. Fantini, S. Loïs, R. Perzynski and V. Peyre,  
\textit{``Colloidal Dispersions of Oxide Nanoparticles in Ionic Liquids : Elucidating the Key Parameters}, ChemRxiv, doi.org/10.26434/chemrxiv.9817208.v1 (2019).

\bibitem{Bacri} J. C. Bacri, R. Perzynski, D. Salin, V. Cabuil, and R. Massart, \textit{``Ionic ferrofluids: A crossing of chemistry and physics''}, \textit{J. Magnetism and Magnetic Materials}, \textbf{17}, 1247 (1981).
\bibitem{Dubois} E. Dubois, V. Cabuil, F. Boué and R. Perzynski, \textit{``Structural analogy between aqueous and oily magnetic fluids''} ,\textit{J. Chem. Phys}. \textbf{111}, 7147 (1999).

\bibitem{Verwey} E.~Verwey and J.~Overbeek, \textit{``Theory of the Stability of Lyophobic Colloids''}, Elsevier, Amsterdam, (1948). 

\bibitem{Dufreche} J. -F.~Dufreche, O.~Bernard, S.~Durand-Vidal et al, \textit{``Frequency-dependent dielectric permittivity of salt-free charged lamellar systems''}, \textit{J. Chem. Phys.}, { \textbf{B 109}, 9873 (2005). }

\bibitem{Durand} S.~Durand-Vidal, M.~Jardat, V.~Dahirel et al, \textit{``Determining the radius and the apparent charge of a micelle from electrical conductivity measurements by using a transport theory: explicit equations for practical use''}, \textit{J. Chem. Phys.},  \textbf{B 110}, 15542 (2006). 

\bibitem{Jardat} . M.~~Jardat, V.~Dahirel, S.~Durand-Vidal et al, \textit{``Effective charges of micellar species obtained from Brownian dynamics simulations and from an analytical transport theory''}, \textit{Molecular Phys.}, \textbf{104}, 3667 (2004). 

\bibitem{heltner}P.~Heltner, Y.~Papir, I.~Krieger, 
\bibitem{heltner}P.~Heltner, Y.~Papir, I.~Krieger,  \textit{``Diffraction of light by nonaqueous ordered suspensions''}, { \textbf{75}}, 1881 (1971).

\bibitem{Kose}A.~Kose, T.~Osake, Y.~Kobayschi et all, \textit{``Direct observation of ordered latex suspension by metallurgical microscope''}, \textit{J.Colloid Interface Sci.} { \textbf{44}}, 330 (1973).

\bibitem{Williams}R.~Williams, R.~Crandall, \textit{``The structure of crystallized suspensions of polystyrene spheres''}, \textit{Phys. Lett.},
{ \textbf{A 48}}, 225 (1974).

\bibitem{Alexander} S.~Alexander, P.~Chaikin, P.~Grant, G.~Morales, P.~Pincus, \textit{``Charge renormalization, osmotic pressure, and bulk modulus of colloidal crystals: Theory''}, \textit{J. Chem.Phys.},  \textbf{B 80}, 5776 (1984)

\bibitem{LLH} L.D.~Landau, E.M.~Lifshitz, \textit{``Fluid Dynamics''}, (Volume 6 of a Course of Theoretical Physics) Elsevier, (2013). 

\bibitem{Jackson} J.D.~Jackson, \textit{``Classical Electrodynamics''} Third Edition, John Wiley \& Sons, Inc, NY (1999).

\bibitem{Dykhne} A.M.~Dykhne, \textit{``Conductivity of a Two-Dimensional Two-Phase System''}, \textit{Soviet JETP},  \textbf{32}, 63, (1971).

\bibitem{O27}L.~Onsager, \textit{ ``On the theory of electrolytes. II''},  \textit{ Physik. Z.} {\textbf{28}}, 277 (1927).

\bibitem{DH}P.~Debye, E.~Huckel, \textit{ ``The theory of the electrolyte II-The border law for electrical conductivity''}, \textit{Physik. Z.} { \textbf{24}}, 305 (1923).
 
\bibitem{LLPK} E.M.~Lifshitz, L.P.~Pitaevski, \textit{``Physical Kinetics''}, (Volume 6 of a Course of Theoretical Physics), Butterworth-Heinenann Ltd., (1981).

\bibitem{GNS} A.~Grosberg, T.~Nguyen, and B.~Shklovskii, \textit{``The physics of charge inversion in chemical and biological systems"} ,  \textit{Rev. Mod. Phys.}  {\textbf{74}}, (2002), 329 (2002).

\bibitem{RS} R.~Robinson, R.~Stokes, \textit{``Electrolyte Solutions''}, Butterworths Scientific Publications, London, (1959).

\bibitem{LG13}L.~Lizana, A.~Grossberg, \textit{``Exact expressions for the mobility and electrophoretic mobility of a weakly charged sphere in a simple electrolyte''}, \textit{Europhysics Letters}  {\textbf{104}}, 68004 (2013).
\end{thebibliography}
\end{document}